\pdfoutput=1

\documentclass[twocolumn]{revtex4}
\usepackage{amssymb,amsmath,graphicx}
\usepackage{color}
\usepackage[T1]{fontenc}
\begin{document}


\title{
  Thermoelectric properties of gapped bilayer graphene
}

\author{Dominik Suszalski}
\affiliation{Marian Smoluchowski Institute of Physics, 
Jagiellonian University, {\L}ojasiewicza 11, PL--30348 Krak\'{o}w, Poland}
\author{Grzegorz Rut}
\affiliation{Marian Smoluchowski Institute of Physics, 
Jagiellonian University, {\L}ojasiewicza 11, PL--30348 Krak\'{o}w, Poland}
\author{Adam Rycerz}
\affiliation{Marian Smoluchowski Institute of Physics, 
Jagiellonian University, {\L}ojasiewicza 11, PL--30348 Krak\'{o}w, Poland}

\date{June 15, 2019}


\begin{abstract}
  Unlike in conventional semiconductors, both the chemical potential and
  the band gap in bilayer graphene (BLG) can be tuned via application of
  external electric field. Among numerous device implications, this property
  also designates BLG as a~candidate for high-performance
  thermoelectric material. In this theoretical study we have calculated
  the Seebeck coefficients for abrupt interface separating weakly- and
  heavily-doped areas in BLG, and for a~more realistic rectangular sample
  of mesoscopic size, contacted by two electrodes. For a~given band gap
  ($\Delta$) and temperature ($T$) the maximal Seebeck coefficient is
  close to the Goldsmid-Sharp value $|S|_{\rm max}^{\rm GS}=\Delta/(2eT)$,
  the deviations can be approximated by the asymptotic expression
  $|S|_{\rm max}^{\rm GS}-|S|_{\rm max}=(k_B/e)\times
  \left[\frac{1}{2}\ln{u}+\ln{}2-\frac{1}{2}+{\cal O}(u^{-1})\right]$,
  with the electron charge $-e$, the Boltzmann constant $k_B$, and
  $u = \Delta/(2k_BT)\gg{}1$.
  Surprisingly, the effects of trigonal warping term in the BLG low-energy
  Hamiltonian are clearly visible at few-Kelvin temperatures, for all accessible
  values of $\Delta\leqslant{}300\,$meV. We also show
  that thermoelectric figure of merit is noticeably enhanced ($ZT>3$)
  when a~rigid substrate suppresses out-of-plane vibrations, reducing the
  contribution from ZA phonons to the thermal conductivity.
\end{abstract}

\maketitle

\section{Introduction}
In recent years, bilayer graphene (BLG) devices made it possible to demonstrate
several intriguing physical phenomena, including the emergence of quantum
spin Hall phase \cite{Qia11,Mah13}, the fractal energy spectrum known as
Hofstadter's butterfly \cite{Dea13,Pon13}, or unconventional superconductivity
\cite{Cao18a,Cao18b}, just to mention a~few. 
From a~bit more practical perspective, a~number of plasmonic and photonic
instruments were designed and build \cite{Yan09,Bon10,Gri12} constituting
platforms for application considerations.
BLG-based thermoelectric devices have also attracted a~significant attention
\cite{Wan11,Chi16,Mah17,Sus18}, next to the devices based on
other two-dimensional (2D) materials \cite{Lee16,Sev17,Qin18}. 

In a~search for high-performance thermoelectric material, one's attention 
usually focusses on enhancing the dimensionless figure of merit
\cite{Mah95,Kim15}
\begin{equation}
\label{ztdef}
  ZT=
  \frac{GS^2T}{K}, 
\end{equation}
where $G$, $S$ and $K$ are (respectively): the electrical conductance,
the Seebeck coefficient quantifying the thermopower, and the thermal
conductance; the last characteristic can be represented as
$K=K_{\rm el}+K_{\rm ph}$, with $K_{\rm el}$ ($K_{\rm ph}$) being the electronic 
(phononic) part.
This is because the maximal energy conversion efficiency is related to $ZT$
via \cite{Iof57}
\begin{equation}
\label{etamax}
  \eta_{\rm max} = \frac{\Delta{}T}{T_h}
  \frac{\sqrt{1+ZT_{\rm av}}-1}{\sqrt{1+ZT_{\rm av}}+\frac{T_c}{T_h}},
\end{equation}
where $T_c$ ($T_h$) is the hot- (or cool) side temperature,
$\Delta{}T=T_h-T_c$, and $T_{\rm av}=(T_c+T_h)/2$.
In particular, for $ZT_{\rm av}=3$ we have
$\eta_{\rm max}>\frac{1}{3}\Delta{}T/T_h$ (with $\Delta{}T/T_h$ 
the Carnot efficiency), and therefore
$ZT>3$ is usually regarded as a~condition for thermoelectric device to be
competitive with other power generation systems. 

As the Seebeck coefficient is squared in Eq.\ (\ref{ztdef}), a~maximum
of $ZT$ --- considered as a~function of the driving parameters specified
below --- is commonly expected to appear close to a~maximum of $|S|$.
Here we show this is not always the case in gapped BLG:
when discussing in-plane thermoelectric transport in a~presence of
perpendicular electric field such that the band gap is much
greater than the energy of thermal excitations ($\Delta\gg{}k_BT$), the maximal
absolute thermopower $|S|_{\rm max}$ corresponds to the electrochemical potential
relatively close to the center of a~gap, namely
$\mu_{\rm max}^{|S|}\approx{}\pm\frac{1}{2}k_BT\ln(2\Delta/k_BT)$,
whereas the maximal figure of merit
$ZT_{\rm max}$ appears near the maximum of the valence band (or the minimum of
the conduction band), i.e., $\mu_{\rm max}^{ZT}\approx{}\pm\Delta/2$. 
In contrast to $ZT_{\rm max}$, $|S|_{\rm max}$ is not directly related to
the value of the transmission probability near the band boundary,  
and these two quantities show strikingly different behaviors
with increasing $\Delta$ for a~given $T$. 

Qualitatively, one can expect that thermoelectric performance of BLG
is enhanced with increasing $\Delta$, since abrupt switching behavior
is predicted for the conductance $G$ when passing $\mu=\pm{}\Delta/2$
for $\Delta\gg{}k_BT$ \cite{Cut69,Rut14b}.
(Moreover, the gap opening suppresses $\kappa_{\rm el}$, reducing the
denominator in Eq.\ (\ref{ztdef}).) Such a~common intuition is build on
the Mott formula, according to which $S$ is proportional to the
logarithmic derivative of $G$ as a~function of the Fermi energy $E_F$.
One cannot, however,  directly apply the Mott formula for gapped systems
at nonzero temperatures, and the link between a~rapid increase of $G(E_F)$
for $E_F\approx{}\Delta/2$ and high $|S|$ is thus not direct in the case
of gapped BLG, resulting in $\mu_{\rm max}^{|S|}\ll{}\Delta/2$. 

The results of earlier numerical work \cite{Hao10} suggest that $|S|_{\rm max}$,
obtained by adjusting $\mu$ for a~given $\Delta$ and $T$, is close to
\begin{equation}
\label{smaxgs}
  |S|_{\rm max}^{\rm GS}=\frac{\Delta}{2eT}, 
\end{equation}
being the Goldsmid-Sharp value for wide-gap semiconductors \cite{Gol99}. 
In this paper, we employ the Landauer-B\"{u}ttiker approach for relatively
large ballistic BLG samples, finding that Eq.\ (\ref{smaxgs}) provides
a~reasonable approximation of the actual $|S|_{\rm max}$ for $\Delta\sim{}k_B{}T$
only. For larger $\Delta$, a~logarithmic correction becomes significant,
and the deviation exceeds $k_B/e$ ($\approx{}86\,\mu$V/K) for
$\Delta\gtrsim{}10\,k_B{}T$.
We further find that---although $|S|_{\rm max}$ grows
monotonically when increasing $\Delta$ at fixed $T$ and may reach
(in principle) arbitrarily large value---$ZT_{\rm max}$ shows a~conditional
maximum at $\Delta_{\star}(T)\sim{}10^2\,k_B{}T$ (for $T\leqslant{}10\,$K). 
An explanation of these findings in terms of a~simplified model for
transmission-energy dependence is provided.

\section{Model and methods}

\begin{figure*}
\centerline{
  \includegraphics[width=0.8\linewidth]{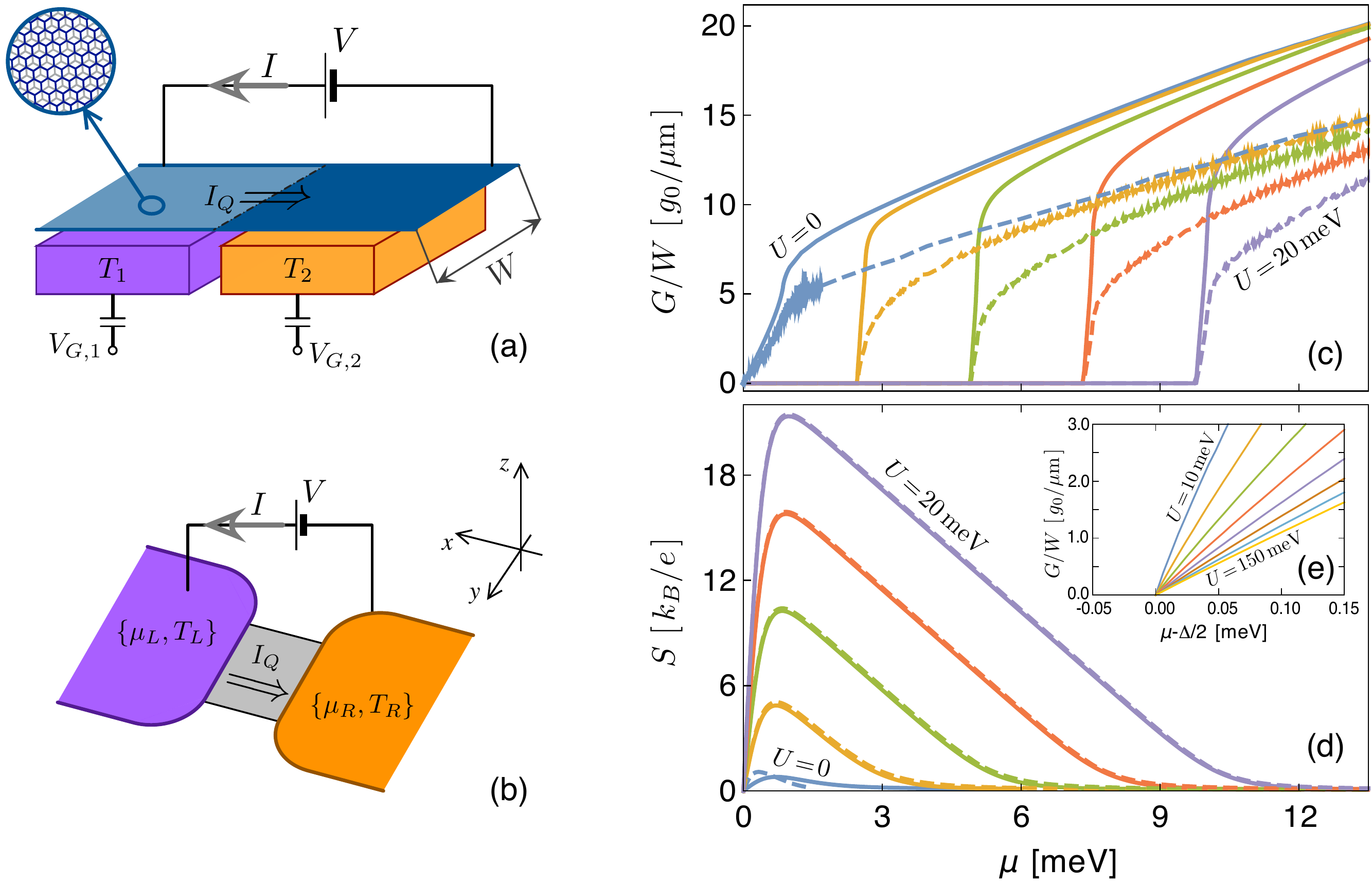}
}
\caption{ \label{gs4panfig} 
  Systems studied numerically in the paper ({\em left\/}) and
  their basic thermoelectric characteristics ({\em right\/}). 
  (a) BLG strip of width $W$ 
  with a~voltage source driving electric current through the strip.
  Two gate electrodes, with different temperatures $T_1$ and $T_2$,
  induce an abrupt interface (dash-dot line) between the weakly-doped
  (light area) and the heavily-doped (dark area) regions.
  (b) A~finite, weakly-doped section of the strip (of length $L$) with leads
  driving both thermal and electric currents. In both cases, additional gate
  electrodes (not shown) allow to tune the electrostatic bias between
  the layers (see Ref.\ \cite{fig1foo}).
  The coordinate system is also shown. 
  (c) Zero-temperature conductance, specified in the units of 
  $g_0=4e^2/h$, and (d) the Seebeck coefficient at $T=5\,$K as functions 
  of the chemical potential. Solid lines correspond to the system of 
  panel (a), dashed lines correspond to the system of panel (b).
  (e) Zoom-in of the conductance as a~function of the chemical potential
  measures from the conduction band minimum ($\mu=\Delta/2$) for the system of
  panel (a). Skew-interlayer hopping is fixed at $t'=0.3\,$eV. 
  The electrostatic bias between the layers is varied from $U=0$ to
  $U=20\,$meV in steps of $5\,$meV (c,d) or from $U=10\,$meV to
  $U=150\,$meV in steps of $20\,$meV (e). (The extreme values of $U$ are
  specified for corresponding lines.)
}
\end{figure*}

The two systems considered are shown schematically in Figs.\ \ref{gs4panfig}(a)
and \ref{gs4panfig}(b).
The first system (hereinafter called an {\em abrupt interface}) clearly
represents an idealized case, as we have supposed that both the doping and
temperature change rapidly on the length-scale much smaller than the Fermi
wavelength for an electron. Therefore, a~comparison with the second system,
in which chemical potentials and temperatures are attributed to macroscopic
reservoirs (the two leads), separated by a~sample area of a~finite length
$L$,  is essential to validate the applicability of our precictions
for real experiments. 

We take the four-band Hamiltonian for BLG \cite{Mac13}
\begin{equation}
  \label{ham1val}
  H=\xi\left(\begin{array}{cccc}
      -U/2 & v_F\pi & \xi{}t_{\perp} & 0\\
      v_F\pi^{\dagger} & -U/2 & 0 & v_3\pi\\
      \xi{}t_{\perp} & 0 & U/2 & v_F\pi^{\dagger}\\
      0 & v_3\pi^{\dagger} & v_F\pi & U/2
    \end{array}\right),
\end{equation}
where the valley index $\xi=1$ for $K$ valley or $\xi=-1$ for $K'$ valley,
$\pi=p_x+ip_y$, $\pi^\dagger=p_x-ip_y$, with $\mbox{\boldmath $p$}=(p_x,p_y)$
the carrier momentum, $v_F=\sqrt{3}at_0/(2\hbar)$ is the Fermi velocity,
$v_3=(t'/t_0)v_F$, $U$ is the electrostatic bias between the layers,
and $a=0.246\,$nm is the lattice parameter. Following Ref.\ \cite{Kuz09},
we set $t_0=3.16\,$eV -- the nearest-neighbor in-plane hopping
energy, $t_\perp=0.381\,$eV -- the direct interlayer hopping energy; 
the skew interlayer hopping energy is set as $t'=0$ or $t'=0.3\,$eV
in order to discuss the role of trigonal warping \cite{fig1foo}.
The band gap $\Delta\approx{}|U|$ for $|U|\ll{}t_\perp$ and $t'=0$ 
(remaining details are given in {\em Supplementary Information, Sec.~I\/}).
Solutions of the subsequent Dirac equation, $H\Psi=E\Psi$, 
with $\Psi=(\Psi_{A1},\Psi_{B1},\Psi_{B2},\Psi_{A2})^T$ the probability amplitudes,
are matched for the interfaces separating weakly- and heavily-doped regions
allowing us to determine the energy-dependent transmission probability $T(E)$
(see {\em Supplementary Information, Sec.~II\/}). 

At zero temperature, 
the relation between physical carrier concentration (the doping) and
the Fermi energy $E_F$ can be approximated by a~close-form expression 
for $t'=0$ and $|E_F|\!-\!\Delta/2\ll{}t_\perp$, namely
\begin{equation}
  \label{nef0appr}
  n(E_F) \approx{}\frac{t_\perp}{\pi{}(\hbar{}v_F)^2}
  {\rm max}\left(0,|E_F|-\Delta/2\right),  
\end{equation}
following from a~piecewise-constant density of states in such a~parameter
range \cite{Mac13}.
The prefactor in Eq.\ (\ref{nef0appr}) $(1/\pi)\,t_\perp/(\hbar{}v_F)^2
\approx 0.268\,$nm$^{-2}$eV$^{-1}$ . 
In a~general situation, it is necessary to perform the numerical integration
of the density of states following from the Hamiltonian $H$ (\ref{ham1val})
(see {\it Supplementary Information, Sec.~I\/}); however, Eq.\ (\ref{nef0appr})
gives a~correct order of magnitude for $\Delta\gg{}k_B{}T$.
In the remaining part of the paper, we discuss thermoelectric characteristics
as functions of the electrochemical potential $\mu$, keeping in mind that
the doping $n=n(|\mu|)$ is a~monotonically increasing
function of $|\mu|$.

Next, we employ the Landauer-B\"{u}ttiker expressions 
for the electrical and thermal currents \cite{Lan57,But85}
\begin{align}
  I &= -\frac{g_sg_ve}{h}\int{}dE\,T(E)\left[f_L(E)\!-\!f_R(E)\right],  \\
  I_Q &= \frac{g_sg_v}{h}\int{}dE\,T(E)\left[f_L(E)\!-\!f_R(E)\right] 
  (E\!-\!\mu), \label{iqland}
\end{align}
where $g_s=g_v=2$ are spin and valley degeneracies,
$f_{L}$ and $f_{R}$ are the distribution 
functions for the left and right reservoirs, with their electrochemical
potentials $\mu_{L}$ and $\mu_{R}$, and  temperatures $T_{L}$ and $T_{R}$.
We further suppose that $\mu_L-\mu_R\equiv{}-eV$ and $T_L-T_R\equiv{}\Delta{}T$
are infinitesimally small (the linear-response regime) and define
$\mu=(\mu_L+\mu_R)/2$ and $T=(T_L+T_R)/2$. The conductance $G$, 
the Seebeck coefficient $S$, and the electronic part of the thermal 
conductance $K_{\rm el}$, are given by \cite{Esf06}
\begin{align}
  G &= \left.\frac{I}{V}\right|_{\Delta{}T=0} = e^2L_0, 
  \label{gland} \\
  S &= -\left.\frac{V}{\Delta{}T}\right|_{I=0} = \frac{L_1}{eTL_0}, 
  \label{sland} \\
  K_{\rm el} &= \left.\frac{I_Q}{\Delta{}T}\right|_{I=0} = 
  \frac{L_0L_2-L_1^2}{TL_0}, 
  \label{keland}
\end{align}
where 
\begin{align}
  \label{llndef}
  L_n&=\frac{g_sg_v}{h}\int{}dE\,T(E)\left(
    -\frac{\partial{}f_{\rm FD}}{\partial{}E}\right)(E-\mu)^n \nonumber\\
    &(n=0,1,2)
\end{align}
with $f_{\rm FD}(\mu,T,E)=
1/\left[\,\exp\left((E\!-\!\mu)/k_BT\right)+1\,\right]$ 
being the Fermi-Dirac distribution function. In particular, for
$T\rightarrow{}0$, Eq.\ (\ref{gland}) reduces to $G=(g_sg_ve^2/h)T(\mu)$,
the well-known zero-temperature Landauer conductance. 

The phononic part of the thermal conductance, occuring in Eq.\ (\ref{ztdef}),
can be calculated using 
\begin{equation}
\label{kaphland}
  K_{\rm ph}=\frac{1}{2\pi}\int{}d\omega\,\hbar{}\omega
  \frac{\partial{}f_{\rm BE}}{\partial{}T}{\cal T}_{\rm ph}(\omega), 
\end{equation}
with $f_{\rm BE}(T,\omega)=
1/\left[\,\exp\left(\hbar\omega/k_B{}T\right)-1\,\right]$
the Bose-Einstein distribution function and ${\cal T}_{\rm ph}(\omega)$
the phononic transmission spectrum.
We calculate ${\cal T}_{\rm ph}(\omega)$ by adopting the procedure developed
by Alofi and Srivastava \cite{Alo13} to the two systems considered in this
work (see {\em Supplementary Information, Sec.~IV\/}).

\section{Numerical results}
Before discussing the thermoelectric properties in details, we present
zero-temperature conductance spectra, which represent the input data
to calculate thermoelectric properties (see Sec.\ II).
Since the Hamiltonian given by Eq.\ (\ref{ham1val}) is particle-hole
symmetric, it is sufficient to limit the discussion to $\mu\geqslant{}0$.  

Typically, the conductance $G(\mu)$ of the finite-strip section, compared
with the case of an~abrupt interface, is reduced by approximately
$50\%$ near the conduction band minimum ($\mu=\Delta/2$) 
due to backscattering on the second interface, and slowly approaches the
abrupt-interface limit for $\mu\gg{}\Delta/2$; see Fig.\ \ref{gs4panfig}(c).
Additionally, for a~rectangular setup of Fig.\ \ref{gs4panfig}(b) oscillations
of the Fabry-P\'{e}rot type are well-pronounces, see Ref.\ \cite{Sus18}.  
In contrast, the Seebeck coefficient is almost identical for both systems,
see Fig.\ \ref{gs4panfig}(d). We further notice that the conductance
near $\mu=\Delta/2$, displayed
in Fig.\ \ref{gs4panfig}(e), is gradually suppressed with increasing $U$. 
(Hereinafter, the bandgap $\Delta$ is determined numerically for the
dispersion relation following from Eq.\ (\ref{ham1val}), see 
{\em Supplementary Information, Sec.~I\/}.
In general, $\Delta<|U|$ \cite{Mac13}).

\begin{figure*}
\centerline{
  \includegraphics[width=0.8\linewidth]{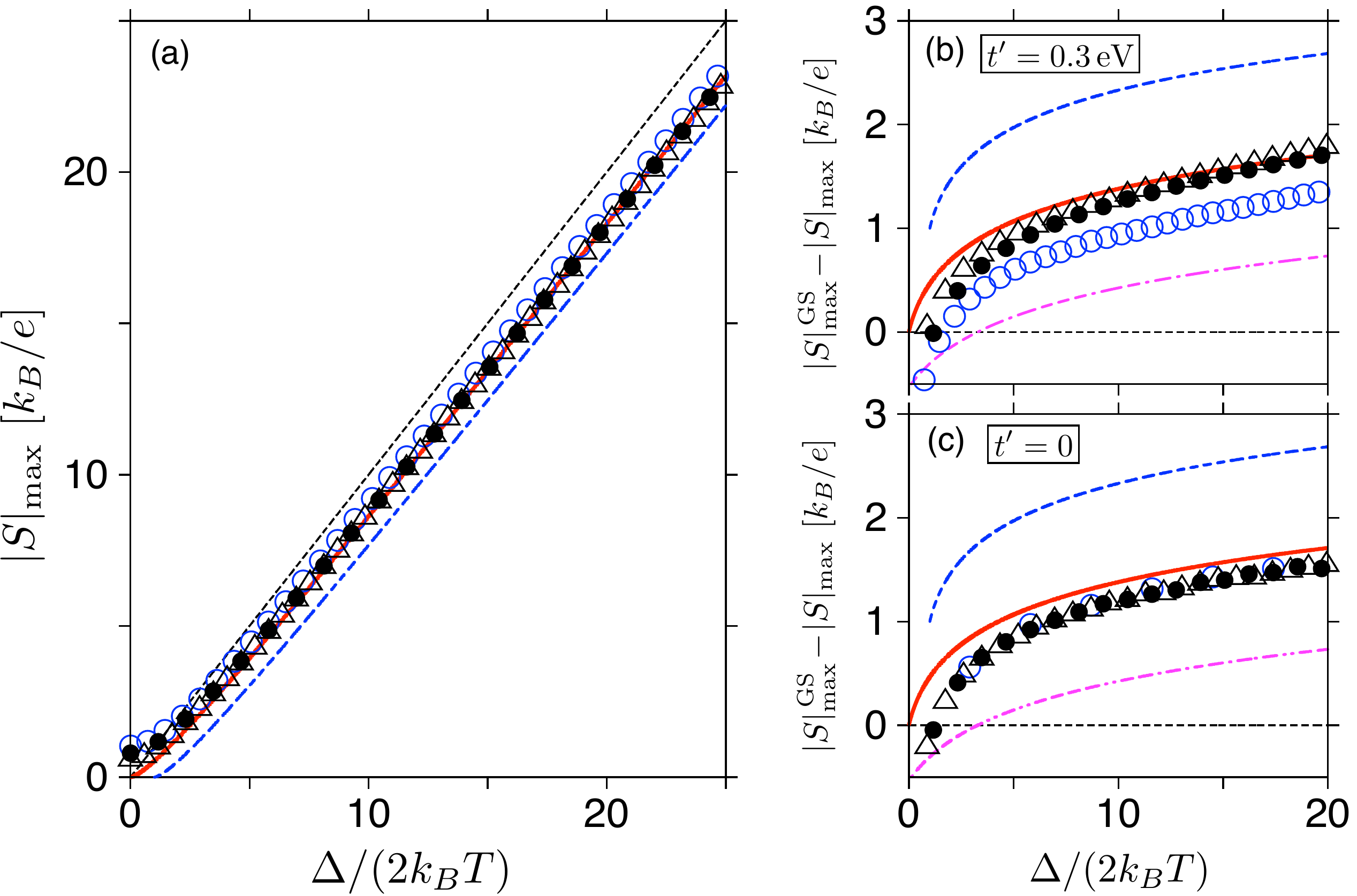}
}
\caption{ \label{smax3panfig} 
  (a) Maximal absolute value of the Seebeck coefficient and
  (b,c) its deviation from the Goldsmid-Sharp value
  $|S|_{\rm max}^{{\rm GS}}=\Delta/(2eT)$ calculated numerically for the system
  of Fig.\ \ref{gs4panfig}(a) (datapoints) as functions of the energy gap  
  for different temperatures $T=1\,$K (open circles), $T=5\,$K (full circles), 
  and $T=10\,$K (triangles).
  Skew-interlayer hopping integral is $t'=0.3\,$eV (a,b) or $t'=0$ (c). 
  Lines depict $S_{\rm max}=S_{\rm max}^{{\rm GS}}$ (black dotted line),
  and predictions of the model for transmission-energy dependence given by
  Eq.\ (\ref{smaxtenm}) with $\alpha=0$ (blue dashed line), $\alpha=1$
  (red solid line), and $\alpha=2$ (magenta dash-dot line --- omitted in panel
  (a) for clarity). 
}
\end{figure*}

The close overlap of the thermopower spectra presented in Fig.\ 
\ref{gs4panfig}(d) allows us to limit the discussion of $|S|_{\rm max}$ 
to the case of an abrupt interface,
see Figs.\ \ref{smax3panfig}(a), \ref{smax3panfig}(b),
and \ref{smax3panfig}(c). 
In order to rationalize the deviations of the numerical data from
$|S|_{\rm max}^{\rm GS}$ given by Eq.\ (\ref{smaxgs}), we propose
a~family of models for the transmission-energy dependence, namely
\begin{equation}
\label{tenmods}
  T^{(\alpha)}={\cal C}(\Delta)\times
    \begin{cases} 
    \delta(E-\frac{1}{2}\Delta)+\delta(E+\frac{1}{2}\Delta)
        & \text{if } \alpha = 0 \\
   \Theta(|E|-\frac{1}{2}\Delta)(|E|-\frac{1}{2}\Delta)^{\alpha-1}
        & \text{if } \alpha > 0
  \end{cases}
\end{equation}
with the prefactor ${\cal C}(\Delta)$ quantifying the transmission propability
near the band boundary
$|E|\approx\frac{1}{2}\Delta$, $\delta(x)$ being the Dirac delta function,
and $\Theta(x)$ being the Heaviside step function.
The analytic expressions presented here, and later in Sec.~IV, are (unless
otherwise specified) valid for any $\alpha\geqslant{}0$, althought 
when comparing model predictions with the numerical data we limit
our considerations to integer $\alpha$. 

In particular, Eq.\ (\ref{tenmods}) leads to
the maximal absolute value of the Seebeck coefficient 
\begin{widetext}
\begin{align}
\label{smaxtenm}
  |S|_{\rm max}^{(\alpha)}\times\left({k_B}/{e}\right)^{-1} &\approx 
  \sqrt{(u+\alpha)(u+\alpha-1)}
  -\ln\left(\sqrt{u+\alpha}+\sqrt{u+\alpha-1}\right)
  \nonumber \\
  &= 
  u-\frac{1}{2}\ln{}\left(4u\right)+\alpha-\frac{1}{2}
  +\frac{3-7\alpha+\alpha^2}{8u}+{\cal O}(u^{-2})
  & \text{for } \alpha\geqslant{}0,
\end{align}
\end{widetext}
where the first asymptotic equality corresponds
to $u=\Delta/(2k_B{}T)\gg{}1$ (see {\it Supplementary Information\/}). 
It is clear from Fig.\ \ref{smax3panfig}(a) that $|S|_{\rm max}^{(\alpha)}$ with
$\alpha=1$ (red solid line) reproduces the actual numerical results (datapoints)
noticeably better than $|S|_{\rm max}^{(\alpha)}$ with $\alpha=0$ (blue dashed line)
or $|S|_{\rm max}^{\rm GS}$  (black dotted line).
What is more, the deviations $|S|_{\rm max}^{\rm GS}-|S|_{\rm max}$, displayed
as functions of $\Delta$, allows one to easily identify the effects of
trigonal warping, see Fig.\ \ref{smax3panfig}(b) for $t'=0.3\,$eV and 
Fig.\ \ref{smax3panfig}(c) for $t'=0$.

\begin{figure*}
\centerline{
  \includegraphics[width=0.6\linewidth]{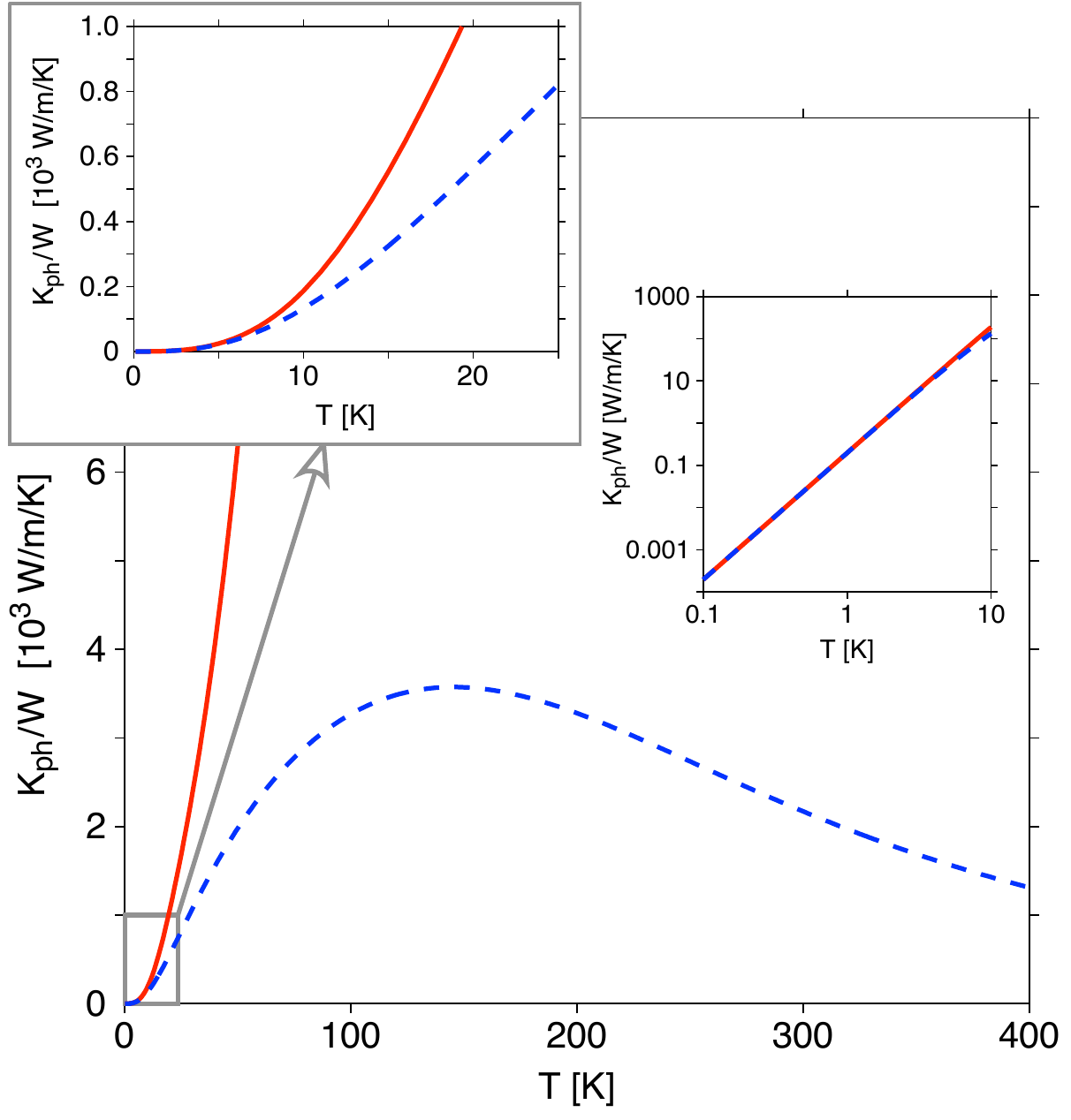}
}
\caption{ \label{kphfig}
  Phononic parts of the thermal conductance for the systems of
  Fig.\ \ref{gs4panfig}(a) (solid line) and Fig.\ \ref{gs4panfig}(b)
  (dashed line) as functions of temperature.
  Insets are zoom in, for low temperatures, in the linear (top left)
  and the log-log (right) scale. 
}
\end{figure*}

\begin{figure*}
\centerline{
  \includegraphics[width=0.9\linewidth]{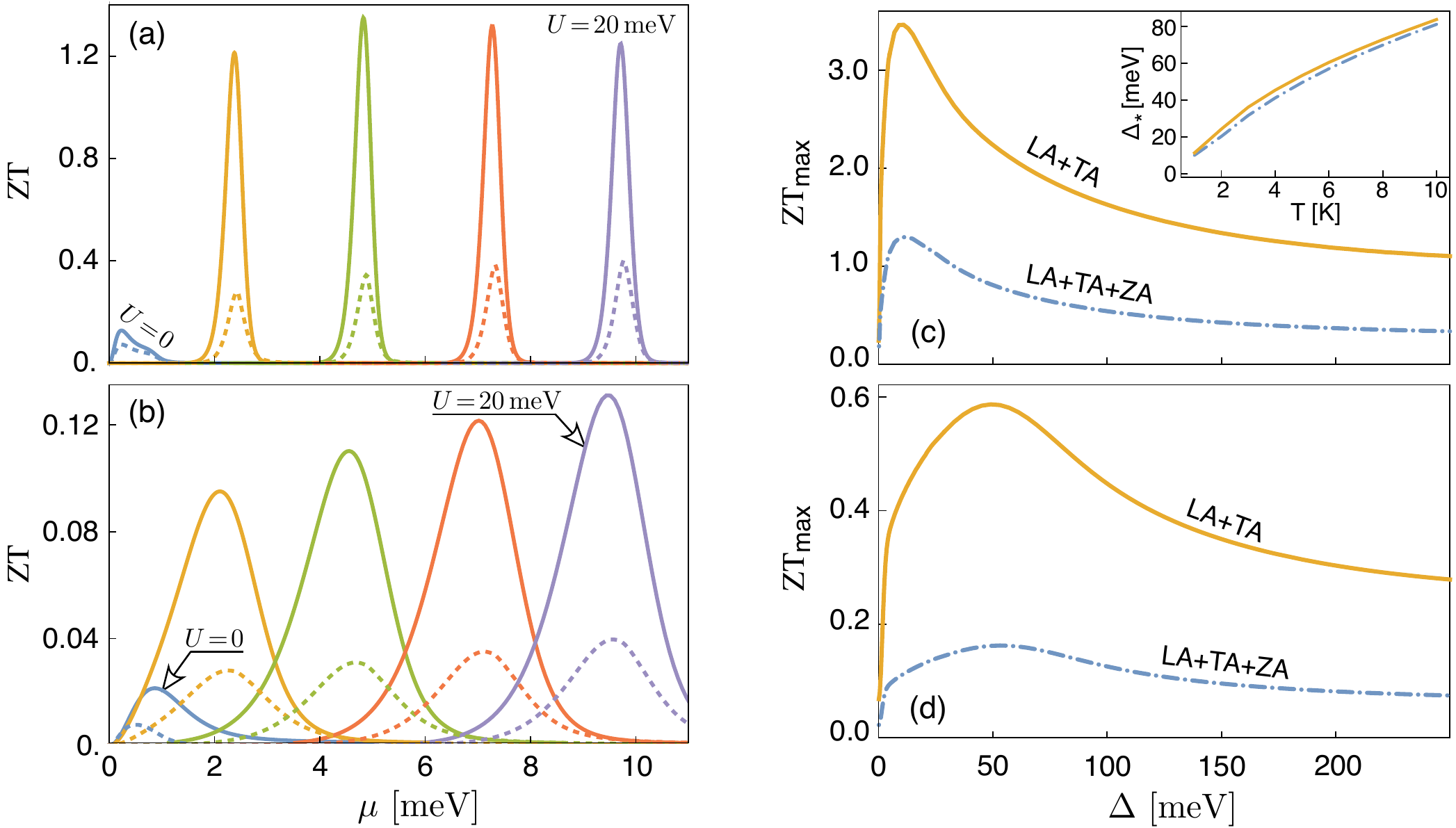}
}
\caption{ \label{ztfig}
  {\em Left:} Thermoelectric figure of merit displayed as a~function of 
  the chemical potential for temperatures (a) $T=1\,$K and (b) $T=5\,$K. 
  Solid lines correspond for the system of Fig.\ \ref{gs4panfig}(a), 
  dashed lines correspond for the system of Fig.\ \ref{gs4panfig}(b). 
  [Remaining parameters are same as in Figs.\ \ref{gs4panfig}(c),
  \ref{gs4panfig}(d).] 
  {\em Right:} Maximal value of the figure of merit for the system of 
  Fig.\ \ref{gs4panfig}(a), versus the energy gap for temperatures 
  (c) $T=1\,$K and (d) $T=5\,$K.
  Inset in panel (c) shows the optimal gap $\Delta_\star$ as a~function
  of temperature. 
  Different lines in panels (c,d) correspond
  to the limit of rigid substrate eliminating ZA phonons (solid lines)
  or the free-standing sample (dashed-dotted lines). 
}
\end{figure*}

Next, we investigate the figure of merit ($ZT$) given by Eq.\ (\ref{ztdef}).
For this purpose, it is necessary to calculate both the electronic part
of thermal conductance ($K_{\rm el}$), that is determined by the
energy-dependent transmission $T(E)$ (see Sec.\ II), as well as
the phononic part ($K_{\rm ph}$), presented in Fig.\ \ref{kphfig}.
We find that for $T<10\,$K the two systems considered show almost equal
$K_{\rm ph}$ ($\propto{}T^3$), and different values of $ZT$
(see Figs.\ \ref{ztfig}(a) and \ref{ztfig}(b)) follow predominantly from
the $G$ reduction discussed above. 
Unlike $|S|_{\rm max}$, a~value of which (for a~given $T$) is limited only by
the largest experimentally-accessible $\Delta\approx{}300\,$meV
\cite{Mac13,Zha09}, $ZT_{\rm max}$ shows well-defined conditional maximum for
the optimal bandgap $\Delta_{\star}/k_B{}T\approx{}100-150$
(at the temperature range $10\,$K$\,\geqslant{}T\geqslant{}1\,$K)
and decreases for $\Delta>\Delta_\star$; see Figs.\ \ref{ztfig}(c)
and \ref{ztfig}(d). 

Also in Figs.\ \ref{ztfig}(c) and \ref{ztfig}(d), we compare $ZT_{\rm max}$
for free-standing BLG, in which all polarizations of phonons (LA, TA, and ZA)
contribute to the thermal conductance (see {\em Supplementary Information,
Sec.~IV\/}) with an idealized case of BLG on a~{\em rigid substrate},
eliminating out-of-plane ($ZA$) phonons.
In the latter case,  $ZT_{\rm max}$ is amplified, approximately by a~factor of
$3$ (for any $\Delta$), exceeding $ZT=3$ for $T=1\,$K and
$\Delta\approx{}\Delta_\star=10\,$meV.

\section{Discussion}
Let us now discuss here why we have identified apparently different behaviors
of $|S|_{\rm max}$ and $ZT_{\rm max}$ with increasing $\Delta$. 
To understand these observations, we refer to the model $T^{(\alpha)}(E)$ 
given by Eq.\ (\ref{tenmods}) with $\alpha\geqslant{}0$, for which
$|S|_{\rm max}^{(\alpha)}$, approximated by Eq.\ (\ref{smaxtenm}), corresponds to 
\begin{align}
  \label{muabsmax}
  \frac{\mu_{\rm max}^{|S|}}{k_B{T}}
  &\approx{}
  \ln\left(\sqrt{u+\alpha}+\sqrt{u+\alpha-1}\right) \nonumber\\
  &\approx{}
  \frac{1}{2}\ln\left(\frac{2\Delta}{k_B{}T}\right)
  \ \ \ \ \text{for } \Delta\gg{}k_B{}T. 
\end{align}
In contrast, the chemical potential corresponding the the maximal $ZT$
is much higher and can be approximated (in the $\Delta\gg{}k_B{}T$ limit)
by 
\begin{equation}
  \label{muztmaxalp1}
  \mu_{\rm max}^{ZT}\approx{}\frac{\Delta}{2}-1.145\,k_B{}T
  \ \ \ \ \text{for } \alpha = 1, 
\end{equation}
or
\begin{equation}
  \label{muztmaxalp2}
  \mu_{\rm max}^{ZT}\approx{}\frac{\Delta}{2}+0.668\,k_B{}T
  \ \ \ \ \text{for } \alpha = 2, 
\end{equation}
where we have further supposed that $K_{\rm ph}\gg{}K_{\rm el}$, being
equivalent to
\begin{equation}
  \label{ztgs2} 
  ZT\approx{}GS^2\frac{T}{K_{\rm ph}(T)}. 
\end{equation}
As the last term in Eq.\ (\ref{ztgs2}) depends only on $T$ 
we can focus now on the {\em power factor} ($GS^2$), a~maximal value of which
can be approximated by
\begin{equation}
  \label{gs2appr}
  \left(GS^2\right)_{\rm max} \approx
  {\cal M}_{\rm max}^{(\alpha)}\frac{g_s{}g_v{}k_B^2}{h}
  \left(k_B{}T\right)^{\alpha-1}{\cal C}(\Delta),
\end{equation}
where the prefactor ${\cal M}_{\rm max}^{(\alpha)}$ depends only on $\alpha$
and is equal to $1.27$ for $\alpha=1$ or to $4.06$ for $\alpha=2$ 
(see also {\it Supplementary Information, Sec.~III\/}).

\begin{figure}
\centerline{
  \includegraphics[width=0.9\linewidth]{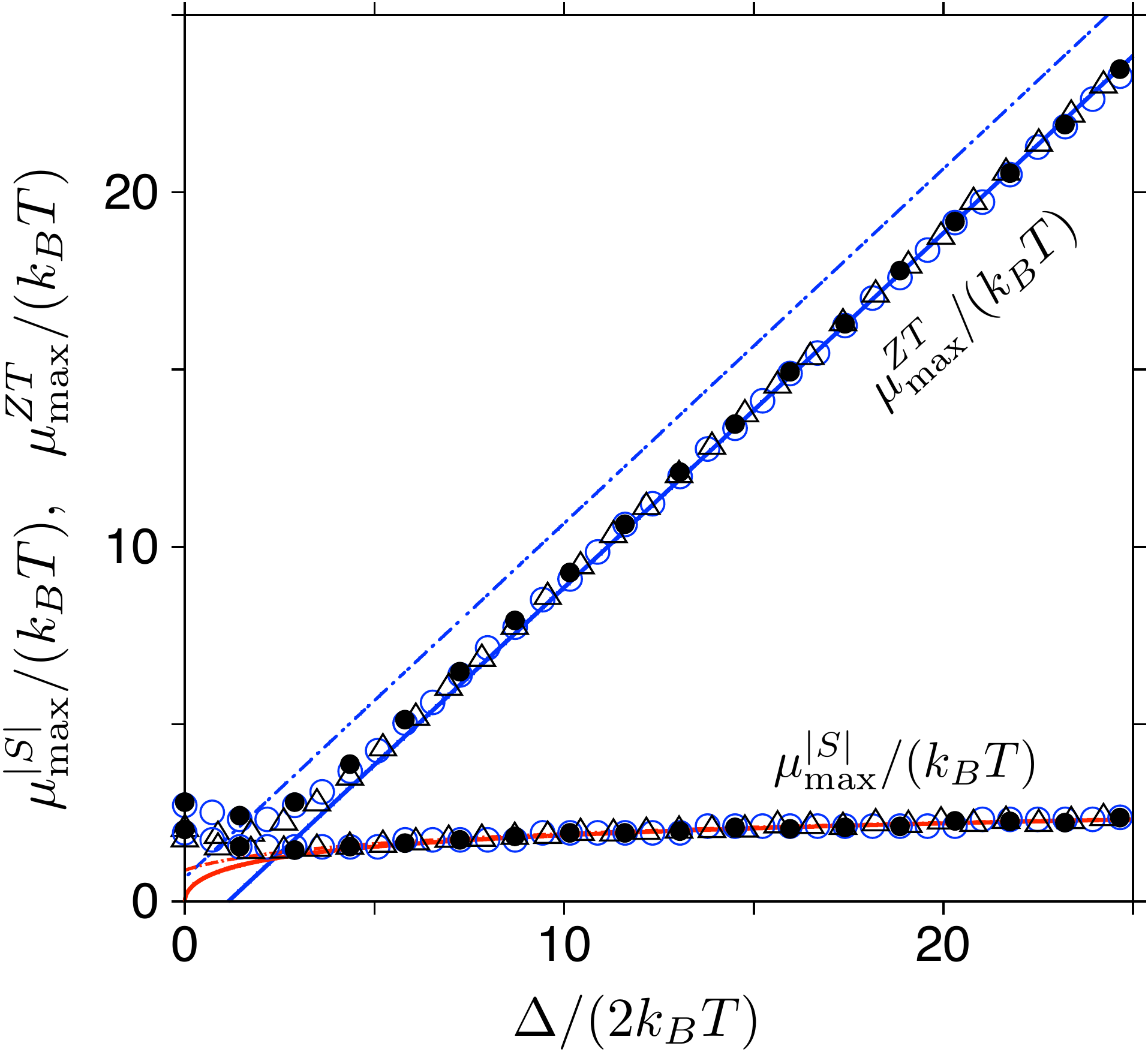}
}
\caption{ \label{muztfig}
  Chemical potential corresponding to the maximal absolute thermopower
  presented in Fig.\ \ref{smax3panfig} ($\mu_{\rm max}^{|S|}$) and the maximal
  figure of merit presented in Fig.\ \ref{ztfig} ($\mu_{\rm max}^{ZT}$)
  as function of the energy gap. 
  Points correspond to same datasets as in Figs.\ \ref{smax3panfig}(a)
  and \ref{smax3panfig}(b) (skew-interlayer hopping integral is $t'=0.3\,$eV). 
  Lines depict approximating Eq.\ (\ref{muabsmax}) with $\alpha=1$ (red solid
  line) and $\alpha=2$ (red dash-dot line),
  Eqs.\ (\ref{muztmaxalp1}) (blue solid line), and (\ref{muztmaxalp2})
  (blue dash-dot line). 
}
\end{figure}

The positions of maxima visible in Figs.\ \ref{ztfig}(a) and \ref{ztfig}(b)
are numerically close to the approximation given by
Eq.\ (\ref{muztmaxalp1}).
Also, the data visualized in Fig.\ \ref{muztfig}, together with the lines
corresponding to Eqs.\ (\ref{muztmaxalp1}) and (\ref{muztmaxalp2}), 
further support our conjucture that the model $T^{(\alpha)}(E)$, given by
Eq.\ (\ref{tenmods}) with $\alpha=1$ (the {\em step-function model\/}),
is capable of reproducing basic
themoelectric characteristics of gapped BLG with a~reasonable accuracy. 

Although it might seem surprising at first glance that the step-function
model ($\alpha=1$) reproduces the actual numerical results much better
than $T^{(\alpha)}(E)$ with $\alpha=2$, as the conductance spectra in
Fig.\ \ref{gs4panfig}(e) exhibit linear, rather than step-line, energy
dependence for $\mu\geqslant{}\Delta/2$. However, for the temperature
range of $1\,$K$\leqslant{}T\leqslant{}10\,$K the $T(E)$ behavior for
$E-\Delta/2\gg{}0.1\,$eV, visualized with the data in Fig.\ \ref{gs4panfig}(c),
preveils [notice that the full width at half maximum
for $-\partial{}f_{\rm FD}/\partial{}E$ in Eq.\ (\ref{llndef}) is
$\approx{}\!3.53\,k_B{}T\,$]. For this reason, a~simple step-function model
grasps the essential features of the actual $T(E)$.

Apart from pointing out that $\mu_{\rm max}^{|S|}\ll{}\mu_{\rm max}^{ZT}\approx{}
\Delta/2$ for $\Delta\gg{}k_B{}T$ (some further implications of this fact
are discussed below), the analysis starting from $T^{(\alpha)}(E)$ models also
leads to the conclusion that --- unlike $|S|_{\rm max}$ that is not directly
related to ${\cal C}(\Delta)$ --- for the figure of merit we have:
$ZT_{\rm max}\propto{}(GS^2)_{\rm max}\propto{}{\cal C}(\Delta)$
[see Eqs.\ (\ref{ztgs2}) and (\ref{gs2appr})].
It becomes clear now that a~striking $ZT_{\rm max}$ suppression for large
$\Delta$ is directly link to the local $G$ suppression for large $U$, 
illustrated in Fig.\ \ref{gs4panfig}(e).
Power-law fits to the datasets presented in Figs.\ \ref{ztfig}(c) and
\ref{ztfig}(d), of the form $ZT_{\rm max}\propto{}\Delta^{-\gamma}$
for $\Delta>\Delta_\star$, lead to $\gamma\approx{}0.5$.
It is worth to stress here that the dispersion relation, and also
the number of open channels as a~function of energy above the
band boundary, $N_{\rm open}(|E|-\Delta/2)$
(see {\em Supplementary Information, Sec.~II\/}), is virtually unaffected by
the increasing $\Delta$. Therefore, the average transmission
for an open channel near the band boundary decreases with $\Delta$. 
This observation can be qualitatively understood by pointing out a~peculiar 
(Mexican hat-like) shape of the dispersion relation for $\Delta>0$ 
\cite{Mac06}. In the energy range $\Delta/2<|E|<|U|/2$ there is a~continuous
crossover from zero transmission (occuring for $|E|<\Delta/2$) to
a~high-transmission range ($|E|>|U|/2$).
As the width of such a~crossover energy range, $(|U|-\Delta)/2$, increases
monotonically with $\Delta$, the continuity of $T(E)$ implies that 
the average transmission near $|E|\approx{}\Delta/2$ decreases with $\Delta$. 

For a~bit more formal explanation, we need to refer the total transmission
probability through an abrupt interface
(see {\em Supplementary Information, Sec.~II\/}). For the incident
wavefunction with the momentum parallel to the barier (conserved during
the scattering) $\hbar{}k_y$ (where $k_y=2\pi{}q/W$ and $q=0,\pm{}1,\pm{}2,\dots$ assuming the periodic boundary conditions) we have
\begin{equation}
  \label{tkyjj}
  T_{k_{y}}=\sum_{m,n}\left|t_{n}^{m}\right|^{2}j_{x}\left(
  \psi_{II}^{n}\right)/j_{x}\left(\psi_{I}^{m}\right), 
\end{equation}
where $\left\{t_n^m\right\}$ is the $2\times{}2$ transmission matrix
($m,n=1,2$ are the subband indices) to be determined via the mode-matching,
and $j_x(\psi_{X}^n)$ is the $x$-component of electric current for the
wave function propagating in the direction of incidence, with $X=I,II$
indicating the side of a~barrier.
For $E=\epsilon{}+\Delta/2$ (with $0<\epsilon\ll{}\Delta/2$)
there are propagating modes in a~weakly-doped area ($X=II$) with $n=1$
(the lower subband) only; in the simplest case without trigonal warping
($t'=0$) we find that the relevant current occuring in
Eq.\ (\ref{tkyjj}) scales as 
\begin{equation}
  j_{x}\left(\psi_{II}^{1}\right) \propto \sqrt{\epsilon/\Delta}. 
\end{equation} 
The above allows us to expect that the full transmission 
scales as $T(E)=\sum_{k_y}T_{k_y}\propto{}\sqrt{\epsilon/\Delta}$ also for 
$t'\neq{}0$, provided that the band gap is sufficiently large 
($\Delta\gg{}E_L$, with $E_L=\frac{1}{4}t_\perp\left(t'/t_0\right)^2$ the 
Lifshitz energy). 
As the number of propagating modes is approximately $\Delta$--independent, 
scaling roughly as $N_{\rm open}\propto{}\sqrt{\epsilon}$, 
we can further predict that zero- (or low-) temperature conductance 
should follow the approximate scaling law
\begin{equation}
  G(\mu)\propto{}\left(\mu-\Delta/2\right)/\sqrt{\Delta}\ \ \ \ 
  (\text{for }\mu\geqslant{}\Delta/2). 
\end{equation}
This expectation is further supported with the numerical data presented
in Fig.\ \ref{gs4panfig}(e). 

An initial increase of $ZT_{\rm max}$ with $\Delta$ for
$0<\Delta\ll{}\Delta_\star$, also apparent in Figs.\ \ref{ztfig}(c)
and \ref{ztfig}(d), can be understood by pointing out that the
electronic and phononic parts of the thermal conductance are of the same
order of magnitude ($K_{\rm el}\sim{}K_{\rm ph}$) in such a~range.
In consequence, an upper bound
to $ZT$ can be written as (up to the order of magnitude)
$ZT\lesssim{}TGS^2/K_{\rm el}$, allowing a~rapid increase of $ZT$
with  $\Delta$ (see {\em Supplementary Information, Sec.~III\/}),
until $K_{\rm el}$ (decreasing with $\Delta$) is overruled by $K_{\rm ph}$
($\Delta$-independent). 

In the remaining part of this section, we briefly discuss the
possible influence of electron-electron interactions, neglected in our
numerical analysis. 

Several experimental works on free-standing BLG report an {\em intrinsic}
(or spontaneous) band gap of $\Delta_{\rm int}(T\!=\!0)\approx{}1.5\,$meV
vanishing above the critical temperature $T_{\rm crit}\approx{}12\,$K
\cite{Yan14,Gru15,Nam16}.
To the contrary, no signatures of an intrinsic band gap are reported for
BLG in van der Waals heterostructures (VDWHs) \cite{Kra18},
in which thermoelectric properties may be significantly enhanced due to
the suppression of out-of-plane (ZA) vibrations.  
Possibly, the above-mentioned difference could be attributed to
a~modification of the effective dielectric constant due to the materials
surrounding a~BLG sample in VDWHs. 
In fact, a~basic mean-field description, relating $\Delta_{\rm int}>0$ to the
alternating spin order, allows one to expect that $\Delta_{\rm int}\sim{}
t_0\exp(-{\rm const.}\times{}t_0/U_{\rm eff})$ (where const.\ $\sim{}1$
is determined by the bandwith), and thus a~moderate decrease of the effective
Hubbard repulsion ($U_{\rm eff}$) strongly suppresses $\Delta_{\rm int}$;
see Ref.\ \cite{Hir85}. 

Although temperatures considered in this paper ($0<T\leqslant{}10\,$K) are
essentially lower then $T_{\rm crit}$, we focus on the case with a~bias between
the layers $|U|\approx{}\Delta\sim{}10-100\,$meV, and much smaller
$\Delta_{\rm int}$ should not affect the physical properties under consideration.

Additionally, the maximal $ZT$ appears near the bottom of the conduction
band or the top of the valence band ($|\mu_{\rm max}^{ZT}|\approx{}\Delta/2$),
where one of the layers is close to the charge-neutrality, and thus one can
expect the Coulomb-drag effects to be insignificant \cite{Gor12}.

\section{Concluding remarks}
We have numerically investigated thermoelectric propeties of large ballistic
samples of electrostatically-gapped bilayer graphene.
A~logarithmic deviation of the maximal absolute thermopower from the
Goldsmid-Sharp relation is identified and rationalized with the help of
the step-function model for the transmission-energy dependence.
In addition to the earlier findings that the trigonal warping term modifies
the density of states \cite{Mac13} and transport properties \cite{Sus18}
also for Fermi energies $\gg{}1\,$meV, we show here that signatures of
trigonal warping may still be visible in thermoelectric characteristics
for the band gaps as large as $\Delta\sim{}100\,$meV. 

Next, the analysis is supplemented by determining the total (i.e., electronic
and phononic) thermal conductance, making it possible to calculate the
dimensionless figure of merit (ZT).
The behavior of maximal $ZT$ with the
increasing gap can also be interpreted in terms of the step-function model,
provided that we supplement the model with the scaling rule for
the typical transmision probability for an open channel near the minimum
of the conduction band (or the maximum of the valence band),
which is $\propto{}\Delta^{-0.5}$ (for large $\Delta$).
This can be attributed to the Mexican-hat like shape of
the dispersion relation. 

Although some other two-dimensional systems with the Mexican-hat like
(or {\em quartic}) \cite{Sev17} dispersion also show enhanced thermoelectric
properties, two unique features of bilayer graphene are worth to stress:
(i) the possibility
of tuning both the chemical potential and the band gap in a~wide range, and
(ii) the ballistic scaling behavior of transport characteristics
with a~barrier width. 
It is also worth to stress that the value of $ZT>1$ (or $ZT>3$ in the absence
of out-of-plane vibrations) is reached at $T=1\,$K for a~moderate bandgap
$\Delta\approx{}10\,$meV, correponding to the electric field of $\approx{}
3\,$mV/\AA, making it possible to consider an experimental setup in which
the spontaneous electric field from a~specially-choosen substrate
(e.g., ferroelectric) is employed to reduce the energy-consumption
in comparison to a~standard dual-gated setup \cite{Zha09}.

High values of $ZT$ at a~$1\,$-Kelvin temperature 
may not have practical device implications {\em per se\/}, however,
we hope that scaling mechanisms identified
in our work will help to find the best thermoelectric among graphene-based
(and related) systems. 
The necessity to reduce the phononic part
of the thermal conductance with a~simultaneous increase of the maximal
accesible band gap (possibly in a~setup not involving an extra power
supply to sustain the perpendicular electric field),
further accompanied by some magnification
of the electric-conductance step on the band boundary, strongly suggests
to focus future studies on graphene-based van der Waals heterostructures.

\section*{Acknowledgments}
We thank Romain Danneau and Bart\l{}omiej Wiendlocha for discussions. 
The work was supported by the National Science Centre of Poland (NCN)
via Grant No.\ 2014/14/E/ST3/00256. Computations were partly performed
using the PL-Grid infrastructure.

\section*{Author contributions}
D.S.\ and A.R.\ developed the code for 
a~rectangular BLG sample, D.S.\ performed the computations for a~rectangular
BLG sample; 
G.R.\ developed the code for the system with abrupt interface
and performed the computations; 
all authors were involved in analyzing the numerical data and writing
the manuscript.

\section*{Additional information}
{\it Supplementary information} accompanies this paper. 





\begin{thebibliography}{}
\bibitem{fig1foo}
Remaining parameters of the systems studied numerically are,
for an abrupt interface of Fig.\ \ref{gs4panfig}(a):
$W=10^3\,l_\perp=1.77\,\mu$m 
(with $l_\perp=\hbar{}v_F/t_\perp$); and for a~rectangular setup of
Fig.\ \ref{gs4panfig}(b): $L=W/20=10^4\,l_\perp$.
\end{thebibliography}
\end{document}